\documentstyle[12pt,twoside,fleqn,espcrc1,fps]{article}

\newcommand{\Z}{Z \!\!\! Z}

\newcommand{\vspsec}{\vspace*{-2mm}}

\newcommand{\AmS}{{\protect\the\textfont2
  A\kern-.1667em\lower.5ex\hbox{M}\kern-.125emS}}

\title{Improved Lattice Actions with Chemical Potential}

\author{W. Bietenholz,
HLRZ c/o Forschungszentrum J\"{u}lich, D-52425 J\"{u}lich, Germany}

\begin{document}
\maketitle

\begin{abstract}
We give a prescription how to include a chemical potential $\mu$
into a general lattice action. This inclusion does not cause any
lattice artifacts. Hence its application to an improved
-- or even perfect -- action at $\mu =0$ yields an improved resp.
perfect action at arbitrary $\mu$. For short-ranged improved actions,
a good scaling behavior holds over a wide region, 
and the upper bound for the baryon density -- which 
is known for the standard lattice actions -- can be exceeded.
\end{abstract}

\section{Introduction}
\vspsec

At chemical potential $\mu \neq 0$, the action of QCD is complex
(see subtitle of this workshop).
With respect to lattice simulations, this means that the standard
Monte Carlo techniques fail. In principle, it is possible to simulate
at $\mu =0$ and include the baryon density $n_{B}$ in the
measured observables by a suitable re-weighting. However, in practice
this is extremely tedious: the ratio of an observable at $\mu \neq
0$ divided by the observable at $\mu =0$ is exponentially suppressed
by the physical volume. Hence the absolute values of the statistical
contributions tend to be many orders of magnitude larger than the
ratio of interest, so that tremendous statistics are required
(``sign problem'') \cite{BG}.
\footnote{In principle, the complex Langevin algorithm is an
alternative \cite{Gau}, but it has not lead to QCD results, again due to the
limited statistics.}

In addition to the statistical error, lattice simulations are
also plagued by systematic errors. The worst source of them are
artifacts due to the finite lattice spacing $a>0$. {\em Improved
lattice actions} are designed to suppress these artifacts,
i.e. the continuum scaling should persist to a good
approximation down to a rather short correlation length in lattice
units, $\xi /a$, or 
up to rather large
values of 
$\mu a$ (below we use lattice units, $a=1$).

As a particularly bad manifestation of lattice artifacts,
there is an upper
bound for $n_{B}$, resp. for the fermion density $n_{f}$.
We will show that improved actions can weaken this
unphysical saturation effect.

In simulations, chiral symmetry appears 
to be restored already at half the pion mass (``onset problem''
\cite{BG,susi}). Usually quenching was 
blamed for that, but attempts to go beyond the quenched
approximation have not really helped in this respect.
The Glasgow group now assumes that this is an effect of low
statistics, but it is also conceivable that lattice
artifacts contribute to this problem. If this is true, then
the use of dynamical fermions together with an improved action
should help to obtain a more continuum-like chiral behavior.

The extreme case of improved actions are {\em perfect actions}:
they eliminate all lattice artifacts. Unfortunately they tend to
involve an infinite number of couplings, hence we can only apply
short-ranged approximations.
A perfect action is constructed by block variable renormalization
group transformations (RGTs) of the type
\begin{equation} \label{trafo}
e^{-S'[\bar \Psi ', \Psi ', U']} = \int D \bar \Psi D \Psi D U
\ e^{-S[ \bar \Psi , \Psi , U]} \
\exp \{ -T[ \bar \Psi ', \bar \Psi , \Psi ' ,\Psi ,U',U] \} \ .
\end{equation}
We start from a fine lattice with action $S[ \bar \Psi , \Psi , U]$,
and introduce a new lattice with a $n$ times coarser lattice spacing,
where the action is given by $S'[\bar \Psi ', \Psi ', U']$ according
to eq. (\ref{trafo}). The transformation term $T$ relates the original
lattice variables to the new ones by averaging in some
way over  the fine lattice sites close to a given coarse site.
It has to provide the invariance of the partition functions,
$Z'=Z$, and of the expectation values, hence of the physical
contents of the theory. The choice for $T$ characterizes the RGT.
After the RGT, the correlation length in lattice units is reduced
by a factor $n$, $\xi' / a' = 1/n \cdot \xi /a$.

Assume a mass $m$ and an inverse temperature $\beta$. 
Now we start from a very fine lattice at mass $m/nN$ and
inverse temperature $\beta nN$, and we perform $N$ RGTs of block
factor $n$. This leads to the desired parameters $m, \ \beta$.
In the limit $nN \to \infty$, this method yields a perfect action
at the desired finite parameters $m, \ \beta$.

\vspsec
\section{Inclusion of the chemical potential $\mu$}
\vspsec

In the continuum, the chemical potential can be included 
by the substitutions
$\bar \psi (-p) \to \bar \psi (-\vec p ,-p_{4}+i\mu )$,
$\psi (p) \to \psi (\vec p ,p_{4}-i\mu )$
in the quark fields, or by replacing $p_{4} \to p_{4} + i\mu$
in the Dirac operator. 
It is not trivial how to apply this substitution to the naive
lattice Dirac operator $i \gamma_{\nu} \sin p_{\nu}+m$.
An early guess, $\sin p_{4} \to \sin p_{4} + i\mu$ does not
have the correct continuum limit. Instead one should
use $\sin (p_{4}+i \mu)$ \cite{HKK}.
Also for the somewhat more complicated Wilson-Dirac operator,
$i \gamma_{\nu} \sin p_{\nu}+m + (r/2) \sum_{\nu} \hat p_{\nu}^{2}$,
$(\hat p_{\nu} = 2 \sin (p_{\nu}/2))$, the substitution $p_{4} \to
p_{4} + i\mu $ works.

If we want to construct  a perfect action, we can use a
standard action (Wilson or staggered)
on the finest lattice and incorporate $\mu$ by this
rule, which we call the {\em standard procedure}. In coordinate
space, it amounts to the substitutions
\begin{equation} \label{stapro}
^{\mu} \bar \Psi (\vec x , t) = e^{\mu t} \bar \Psi (\vec x ,t) \ ,
\quad ^{\mu} \Psi (\vec x , t) = e^{-\mu t} \Psi (\vec x ,t) \ .
\end{equation}
Hence $\mu$ is treated on the lattice consistently as an imaginary
constant Abelian gauge potential $A_{4}=i\mu$.
Therefore, $^{\mu}\bar \Psi$ and $^{\mu}\Psi$ are ``parallel
transported to $t=0$'', which confirms the gauge invariance
of this procedure. This observation holds for all sorts of lattice
actions, hence the standard procedure (\ref{stapro}) can always be
applied.

We preserve this useful property also under the RGT by using
the transformation term
\begin{equation} \label{trafo2}
T[ ^{n\mu}\bar \Psi ', ^{\mu}\bar \Psi,^{n\mu}\Psi ',^{\mu}\Psi ,
U' ,U] , \
^{n\mu}\bar \Psi '(\vec x' ,t') = e^{\mu n t'}\bar \Psi '
(\vec x' , t'), \
^{n\mu} \Psi '(\vec x' ,t') = e^{-\mu n t'} \Psi ' (\vec x' , t').
\end{equation}
If $t \in [0 , \beta]$, then $t' \in [0, \beta /n]$, and since
distances are measured after the RGT in coarse lattice units,
the transportation distance to $t=t'=0$ is divided by $n$.
This is compensated by substituting $n \mu$ in the coarse fields.

The result of the RGT is $S' [ ^{n\mu}\bar \Psi ',^{n \mu}
\Psi ',U']$, and it is easy to see that this is {\em identical}
to the following construction:
first let $\mu=0$ and perform the RGT to arrive at 
$S' [\bar \Psi ', \Psi ',U']$. Now we include $\mu$ in the
blocked action again by the standard procedure (\ref{stapro})
(where $\mu$ is multiplied by $n$ due to the rescaling to the new
lattice units).

Now the RGT can be iterated. In order to arrive at a chemical
potential $\mu$ in the final action, we start with $\mu /nN$
on the finest lattice, and perform $N$ RGTs of block factor $n$,
in analogy to the mass.
The limit $nN\to \infty$ yields a perfect action with $\mu$.

From the analysis of one RGT we conclude that we can switch off $\mu$
in the beginning, iterate the RGT all the way to the perfect
action, and then include $\mu$ again by the procedure
(\ref{stapro}). This leads to a perfect action at chemical
potential $\mu$ \cite{chempap}.

{\em The standard procedure to include $\mu$ preserves perfectness,
it does not cause any lattice artifacts and it is therefore
perfect itself.} We emphasize that this result applies to
any perfect action for the fully interacting quantum theory.

In practice, such actions are hard to construct
(for recent reviews, see \cite{perfrev}). What is more
realistic is a {\em classically perfect} approximation, where
the functional integral in eq. (\ref{trafo}) is simplified to
a minimization of the exponent on the right-hand side.
A number of 2d studies suggest that classically
perfect actions also suppress the lattice artifacts very strongly.
Consideration of the transformation term shows that including
$\mu$ in a classically perfect action by the procedure
(\ref{stapro}) leads again to a classically perfect action with $\mu$.

An alternative improvement program for lattice actions starts from
a standard action and tries to eliminate the artifacts order by
order in the lattice spacing ({\em Symanzik's program}).
Assume that this is realized to some order at $\mu =0$,
\footnote{What has been achieved -- to a good accuracy -- is the
$O(a)$ improvement of QCD with Wilson fermions \cite{Lat97}.
However, going beyond that is not realistic in the foreseeable
future.}
and we include $\mu$
once more by the standard procedure. The above properties
suggest that we obtain an action with $\mu$, which is still free of
lattice artifacts to the same order as it was the case at $\mu =0$.

\vspsec
\section{Scaling behavior of a free fermion gas as an example}
\vspsec

A perfect lattice action for free fermions with mass $m$ at $\mu =0$
reads \cite{QuaGlu}
\begin{equation}
S^{*}[\bar \Psi , \Psi ] = \int_{-\pi}^{\pi} \frac{d^{4}p}{(2\pi )
^{4}} \bar \Psi (-p) \Delta^{-1}(p) \Psi (p) \ , \quad
\Delta (p) = \sum_{l \in \Z^{4}} \frac{\prod_{\sigma =1}^{d}
\hat p_{\sigma}^{2}/(p_{\sigma}+2\pi l_{\sigma})^{2}}
{i \gamma_{\nu}(p_{\nu}+2\pi l_{\nu})+m} + \frac{1}{\alpha} \ ,
\end{equation}
where $\alpha$ is a free RGT parameter.
At $m=0,\ \alpha = \infty$ the action is chirally symmetric but
non-local. For any finite $\alpha$ the action is local:
its couplings in coordinate space decay exponentially.
Since the limit $nN\to \infty$ relates the system directly
to the continuum, there is still chiral symmetry {\em in the 
observables} \cite{Schwing}. Furthermore, the action itself also
keeps a remnant continuous chiral symmetry \cite{ML}.

For applications, it is important to optimize the locality,
so that the truncation of the couplings is not too harmful.
This is achieved by choosing \ $ \alpha = (e^{m}-m-1)/m^{2}$
\cite{QuaGlu}.

According to section 2, the corresponding perfect action at finite
$\mu$ is characterize by $\Delta (\vec p ,p_{4}+i\mu )$.
A successful and yet applicable approximation is a truncation to
the couplings in a unit hypercube by means of periodic boundary
conditions.
The corresponding couplings at $\mu =0$ for various masses
are given in Ref. \cite{Lat96}. 

The RGT used for the above perfect action for Wilson-type fermions
is based on the usual block average (BA) scheme. If one constructs
the analogous perfect action for staggered fermions \cite{stag1},
then the truncation to the same number of degrees of freedom
as before (couplings distance components $\pm 1, \pm 3$
resp. $0,\pm 2$ at $m=0$) is not satisfactory. The locality -- and
hence the quality after truncation -- can be improved significantly
by using  instead a blocking scheme that we call ``partial
decimation'' (PD) \cite{stag2}.

We now focus on $m=0$ and temperature $T=0$.
The pressure and baryon density 
of the free fermion gas are given by
\begin{eqnarray} \nonumber
P &=& \frac{\mu^{4}}{6 \pi^{2}} \ , \quad n_{B} =
\frac{2\mu^{3}}{9 \pi^{2}} \qquad 
{\rm in~the~continuum,~and~obtained~from} \\
P &=& \int_{-\pi}^{\pi} \frac{d^{4}p}{(2\pi )^{4}}
\log \frac{{\rm det} \Delta (\vec p ,p_{4})}
{{\rm det} \Delta (\vec p ,p_{4}+i\mu)} \ , \quad 
n_{B} = \frac{1}{3} \frac{\partial}{\partial \mu} P
\qquad {\rm on~the~lattice.}
\end{eqnarray}
For comparison, we insert a number of lattice fermion propagators 
$\Delta$. For Wilson-type fermions, we probe the Wilson fermion with Wilson
parameter $r=1$, a Symanzik improved version thereof called D234
with additional couplings on the axes \cite{D234}, and the truncated perfect 
{\em ``hypercube fermion''} (HF). 
For staggered fermions we insert the staggered
standard action, the Naik fermion \cite{Naik}
(which is Symanzik improved in
the same way as the D234 fermion), and again the truncated perfect
actions for the BA and the PD scheme.
Fig. 1 shows the scaling ratios $P/\mu^{4}$ and $n_{B}/\mu^{3}$
for this set of fermions. Of course, they all reproduce the
correct continuum values at $\mu \to 0$. As $\mu$ increases,
the standard actions deviate very soon, the Symanzik improved actions
do well up to a certain $\mu$ and then collapse
completely, but the truncated perfect actions keep close to the
continuum value up to remarkably large $\mu$.
Their scaling region is extended by one order of magnitude
compared to the standard actions.

One could also consider $\chi_{B}/\mu^{2}$, where $\chi_{B}=
\frac{\partial}{\partial \mu} n_{B}$ is the baryon number
susceptibility, but in all these cases the qualitative
behavior is very similar, and it is also in agreement with 
the scaling of $P/T^{4}$
at $\mu =0$ \cite{Lat96,stag2}, hence it is manifestly systematic.
\begin{figure}[hbt]
   \begin{tabular}{cc}
      \hspace{-0.6cm}
\def\fpsangle{270} \epsfxsize=57mm \fpsbox{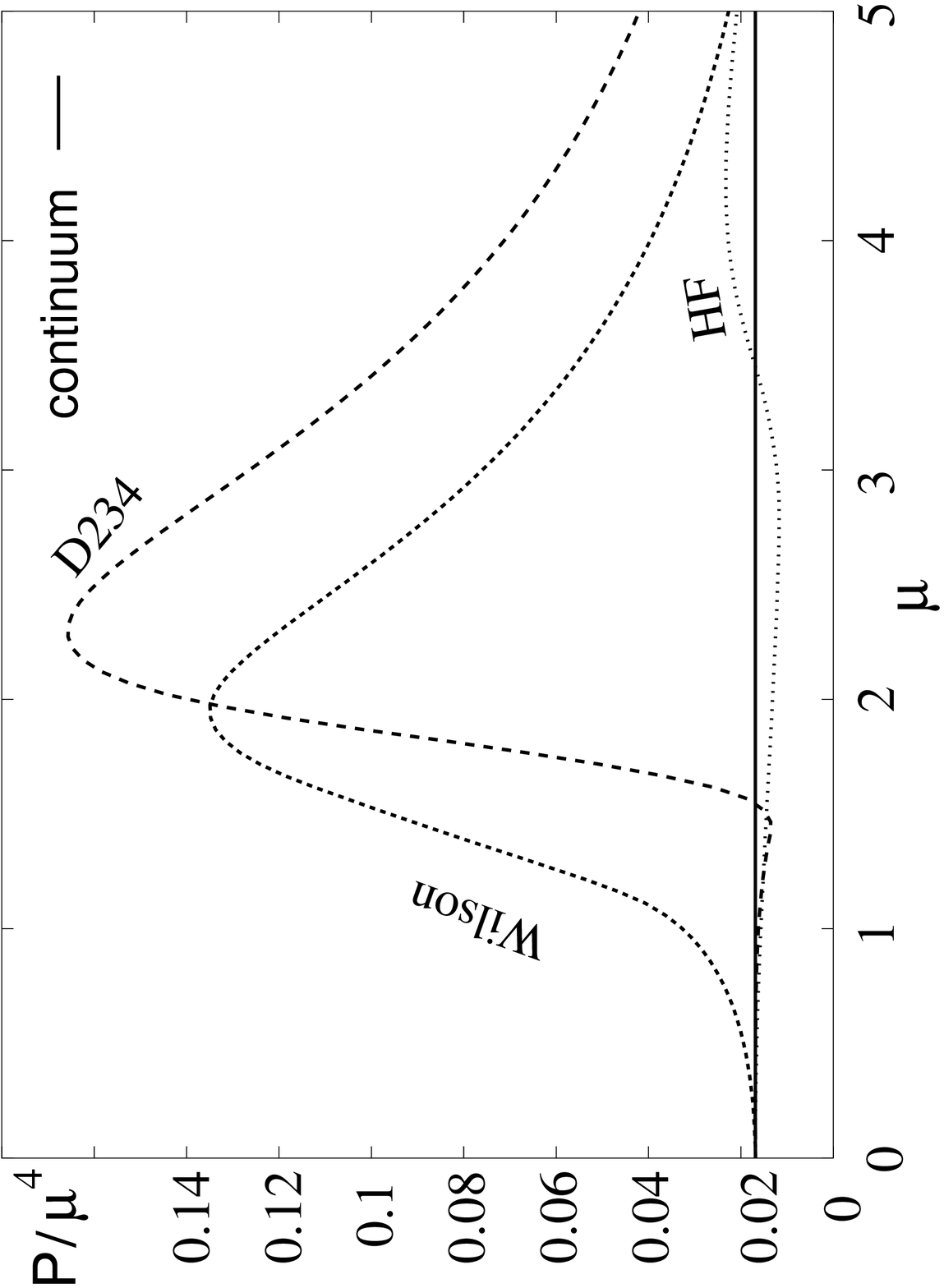} &
     \hspace{-7mm}
\def\fpsangle{270} \epsfxsize=57mm \fpsbox{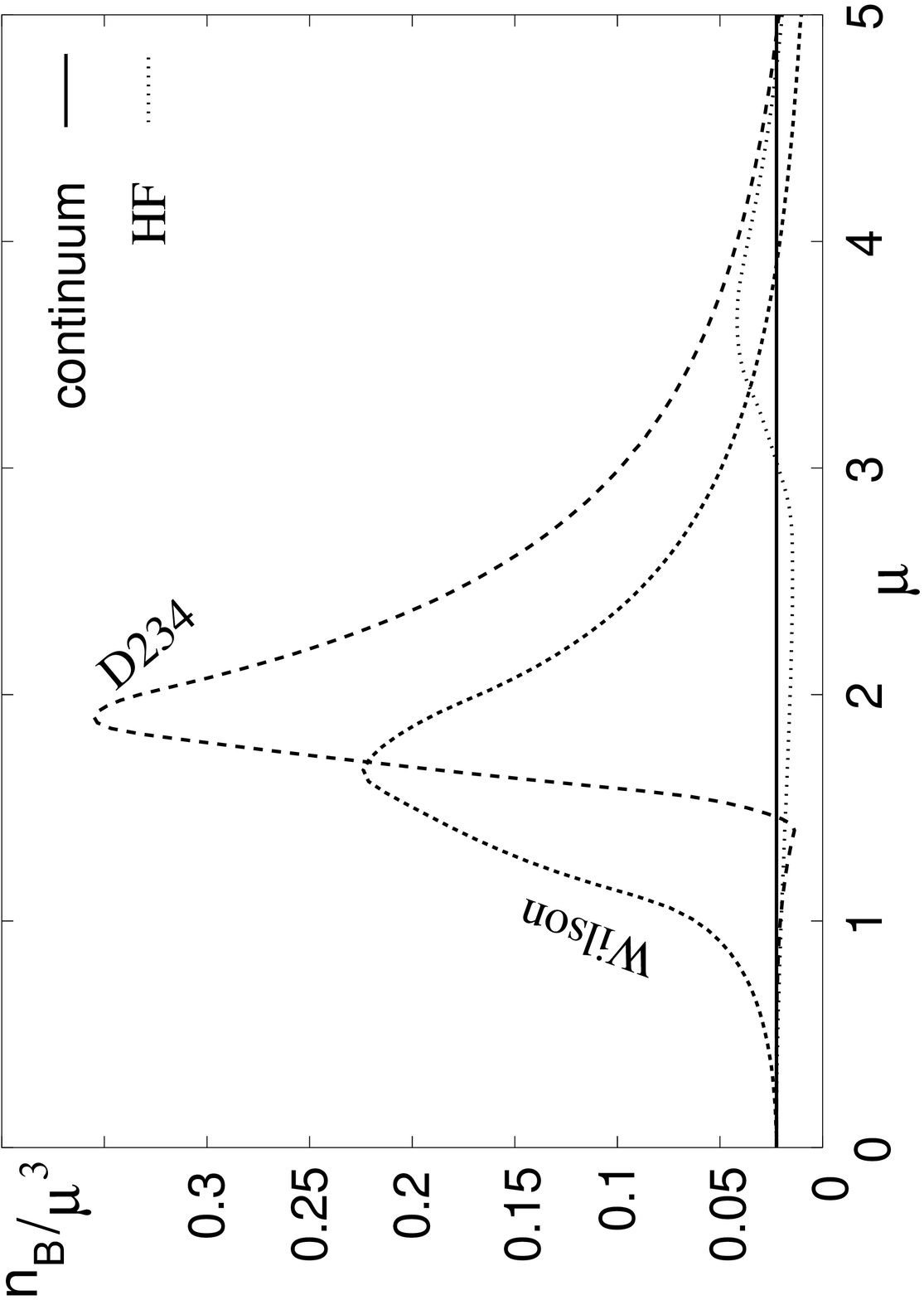}
   \end{tabular}
   \begin{tabular}{cc}
      \hspace{-0.6cm}
\def\fpsangle{270} \epsfxsize=57mm \fpsbox{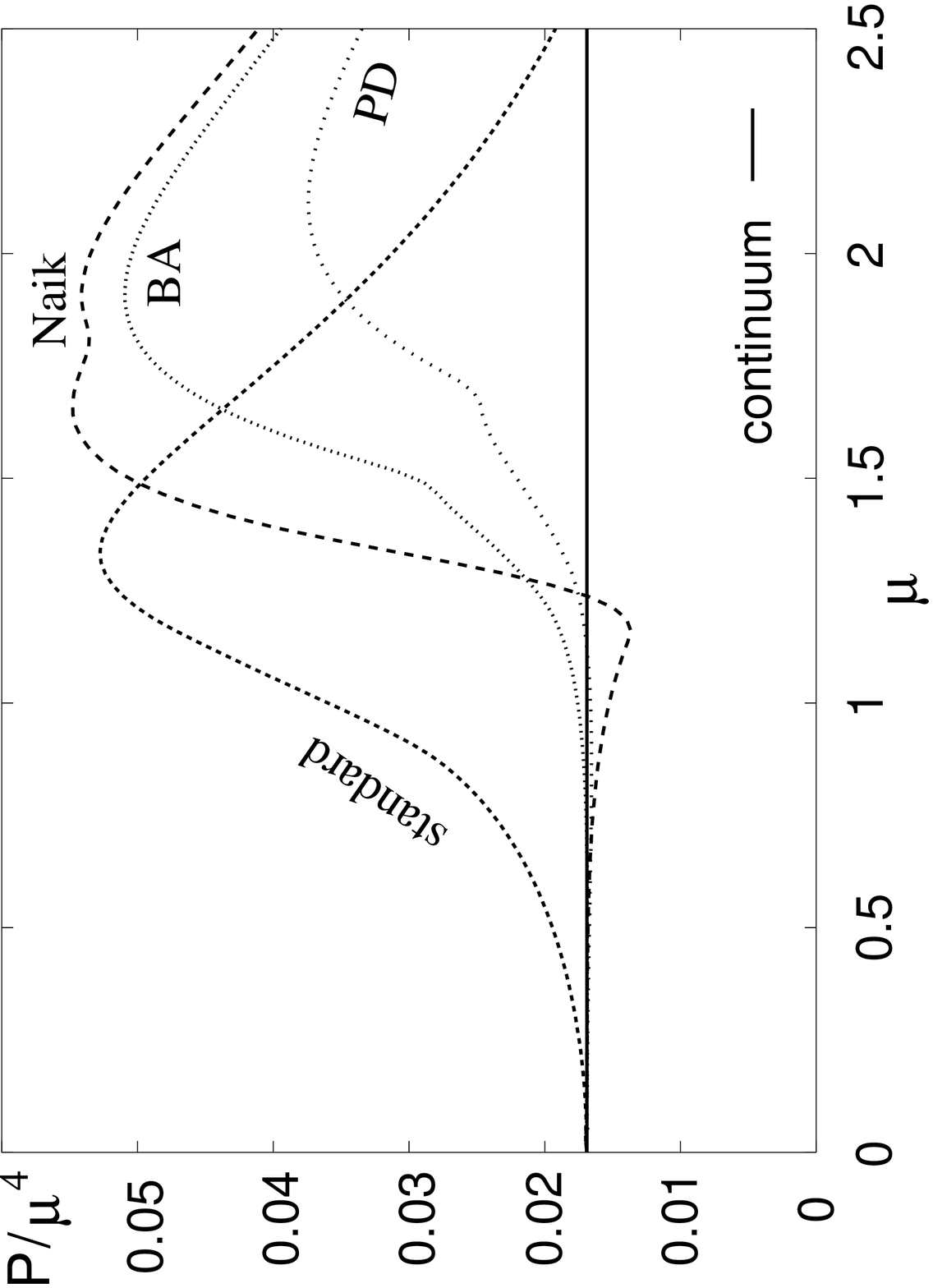} &
      \hspace{-7mm}
\def\fpsangle{270} \epsfxsize=57mm \fpsbox{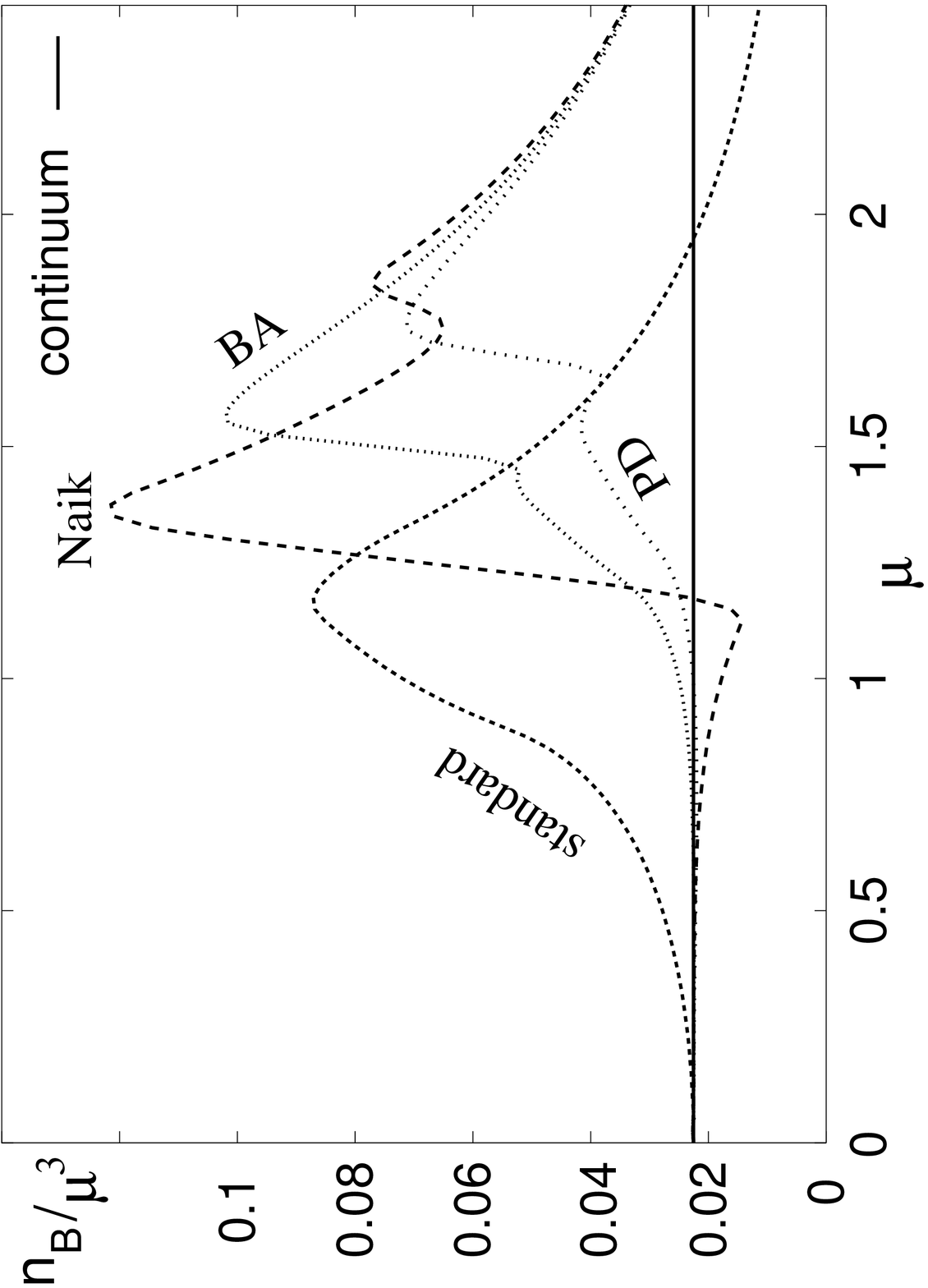}
   \end{tabular}
   \vspace{-8mm}
   \caption{{\it The scaling of Wilson-type fermions (on top)
and staggered fermions (below), both in a number of variants
discussed in the text. The D234 and the Naik fermion are Symanzik
improved, and the fermions denoted as HF, BA and PD are 
truncated perfect.}}
   \label{scal}
\vspace{-8mm}
\end{figure}


\vspsec
\section{The fermion density}
\vspsec

We now consider the fermion density $n_{f}$.
\footnote{At this point I thank especially M.-P. Lombardo, who
suggested to me to consider also $n_{f}(\mu )$ itself.}
It is well known that the standard lattice
actions suffer from an upper bound for $n_{f}$. 
We choose the normalization such
that naive fermions saturate
at $n_{f} =32$. This corresponds to an occupation of each site
by $2^{d}$ species, each with spin up and down.

For Wilson fermions the doublers are
heavier, hence $\mu$ must be larger to exceed their masses,
and the saturation is delayed. However, for $r\neq 1$ this is 
unimportant in the limit $\mu \to \infty$, so we have
again $n_{f,max}=32$.
The case of $r=1$ is special: here the mass of the time-like
doublers diverges, hence $n_{f,max}=16$.

From Pauli's principle one might be tempted
to conclude that $n_{f}=32$ is an absolute upper bound for all
lattice actions. However, since this bound is a lattice
artifact, this leads to a puzzle about the behavior of a perfect
action. The solution is that the above argument is wrong. Considering
\begin{equation}
S [\bar \Psi ,\Psi ] = \sum_{\vec x,t} \sum_{\vec y,s}\ 
\bar \Psi_{\vec x,t} \ e^{\mu (t-s)} \
\Delta^{-1}(\vec x -\vec y ,t-s) \ \Psi_{\vec y,s} \ ,
\end{equation}
we recognize that the exponential growth of the factor containing $\mu$
is limited by the maximal coupling distance in the temporal
direction, which occurs in $\Delta^{-1}$. 
(In fact, such a saturation limit also exists for bosons.)
For the perfect action this distance is infinite,
hence $n_{f}$ has no upper bound, in
agreement with the absence of artifacts. To turn it the other way
round, we can conclude that (in infinite volume)
no perfect action with couplings only in a finite range can exist,
and this is indeed correct.
\footnote{The only exception for matter fields
is the case $d=1$, which does, however, not lead to a contradiction.}
For the hypercube fermion we have again 
$(t-s)_{max} =1 \ \to \ n_{f,max}=32$,
but the saturation is significantly delayed.
The D234 action has $(t-s)_{max} =2$, so
one could expect $n_{f,max}=64$. But there
is a cancelation going on, similar to the Wilson fermion
at $r=1$, which reduces $n_{f,max}$ again to 32. If
we keep that structure and vary the couplings, then $n_{f}$ does 
rise up to 64. But in both cases, the saturation occurs at
practically the same value of $\mu$ as it happens for the
$r=1$ Wilson fermion. This behavior is shown in Fig. 2,
which also illustrates the Wilson fermion at various values of $r$.
For instance at $r=2$, $n_{f}(\mu )$ rises quite steadily to 32.
If $r$ decreases towards 1, then half of the particles
in the spectrum turn very heavy. The curve reaches a first plateau
at 16, and when $\mu$ catches up with this heavy species, it performs
a second jump up to 32. This looks similar if we approach $r=1$
from below. If $r$ is exactly 1, then the plateau at 16 extends to
infinity.
\begin{figure}[hbt]
   \begin{tabular}{cc}
      \hspace{-0.6cm}
\def\fpsangle{270} \epsfxsize=57mm \fpsbox{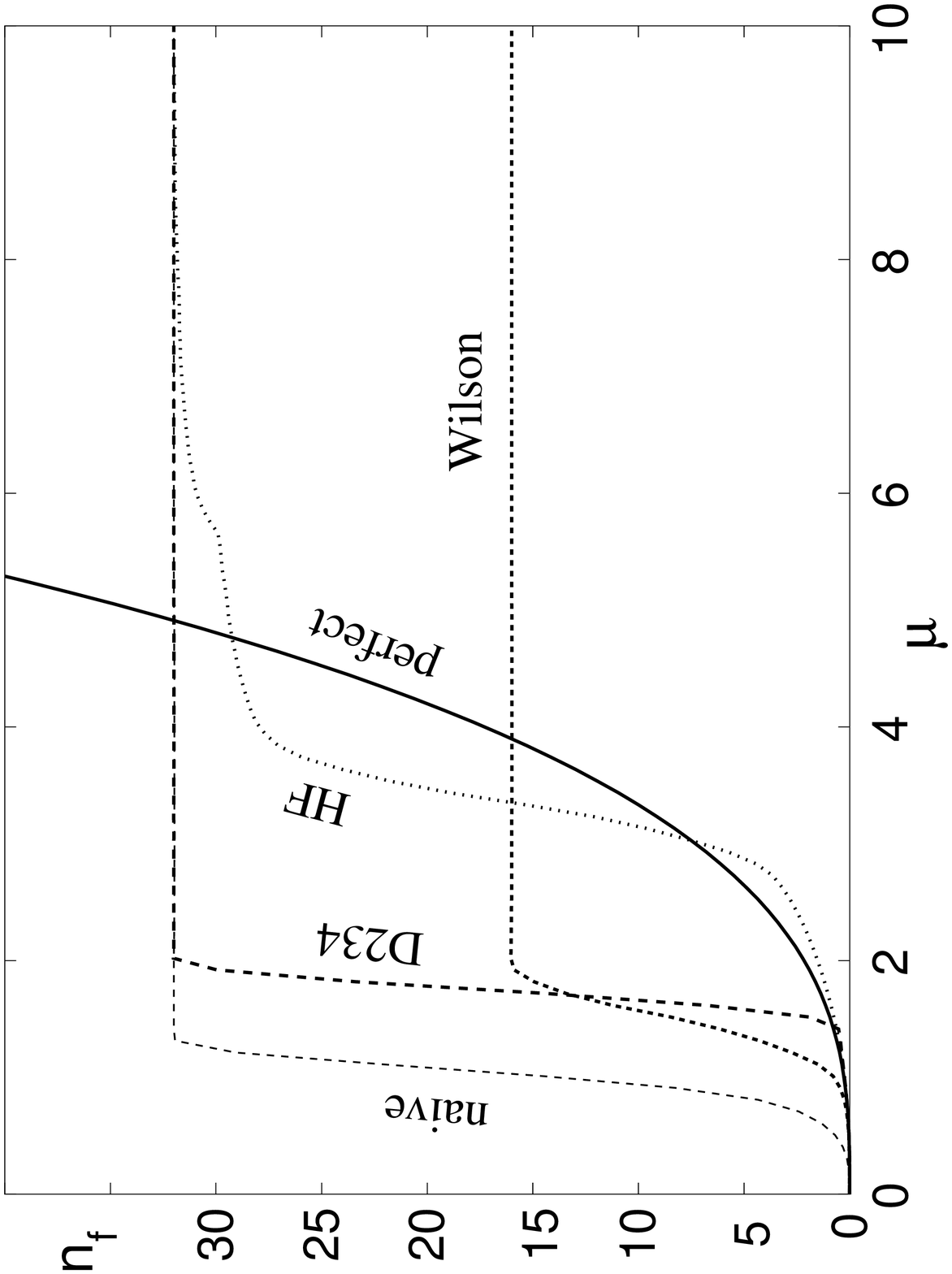} &
      \hspace{-6mm}
\def\fpsangle{270} \epsfxsize=57mm \fpsbox{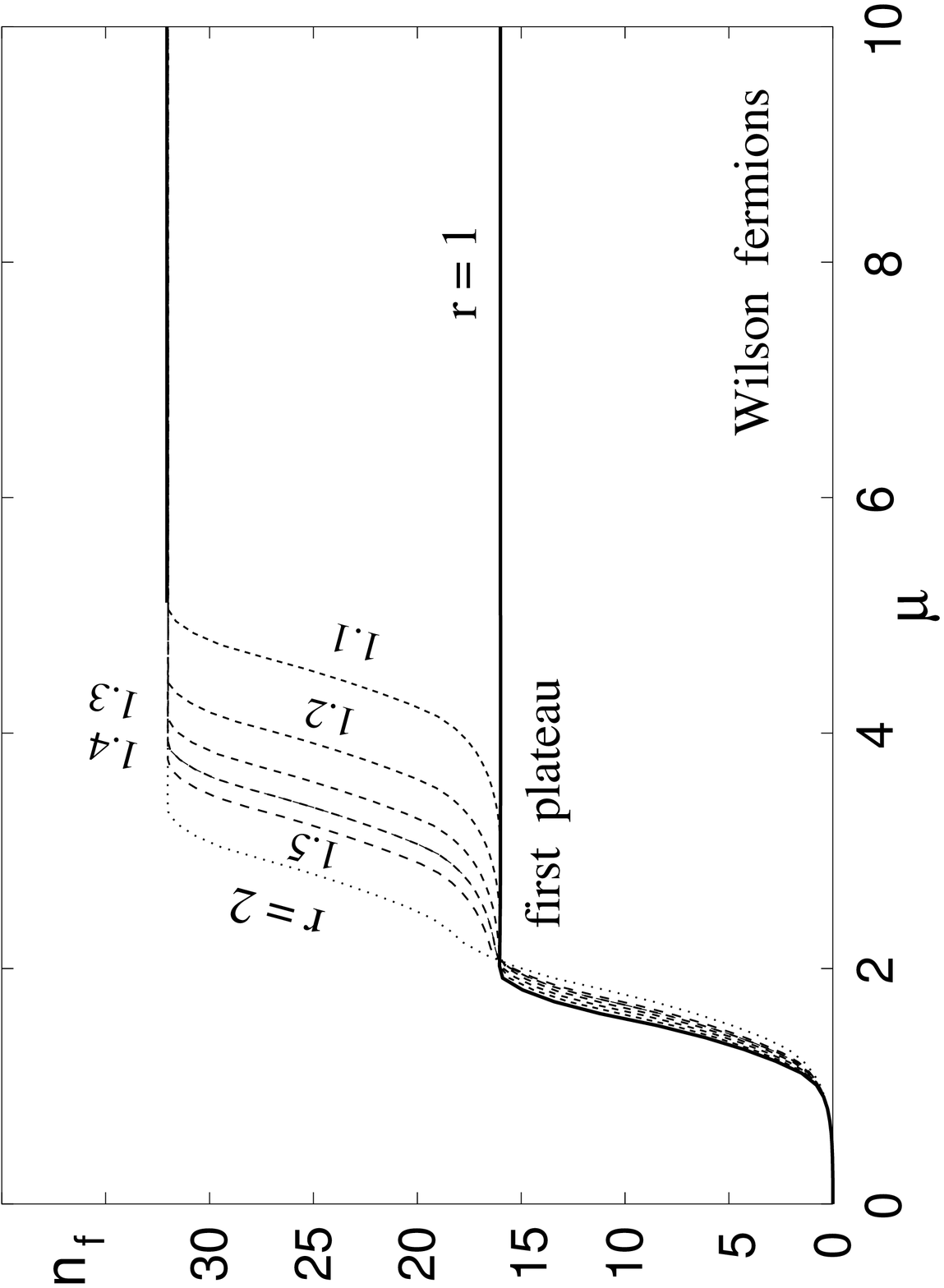}
   \end{tabular}
   \begin{tabular}{cc}
      \hspace{-0.6cm}
\def\fpsangle{270} \epsfxsize=57mm \fpsbox{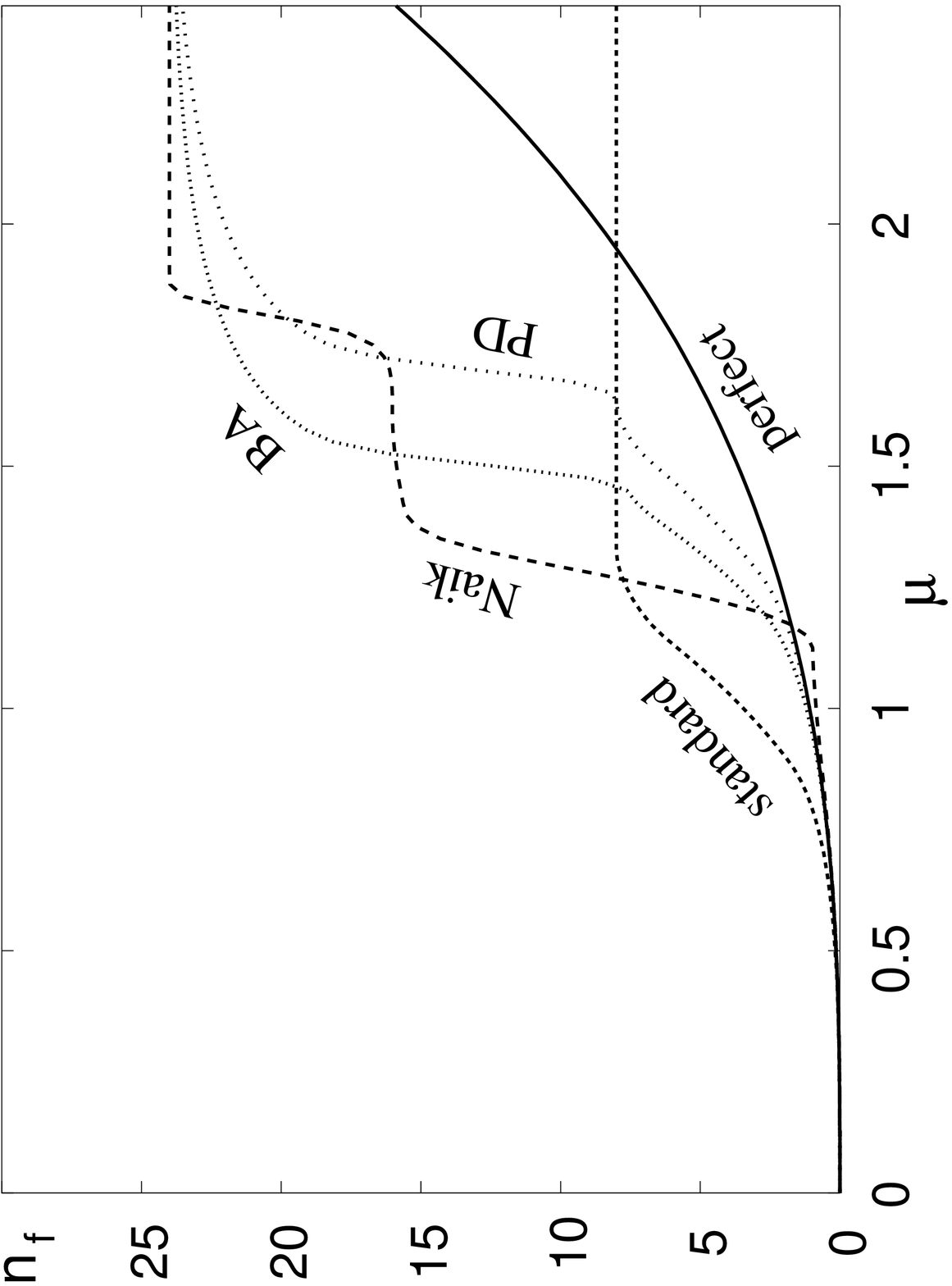} &
      \hspace{-6mm}
\def\fpsangle{270} \epsfxsize=57mm \fpsbox{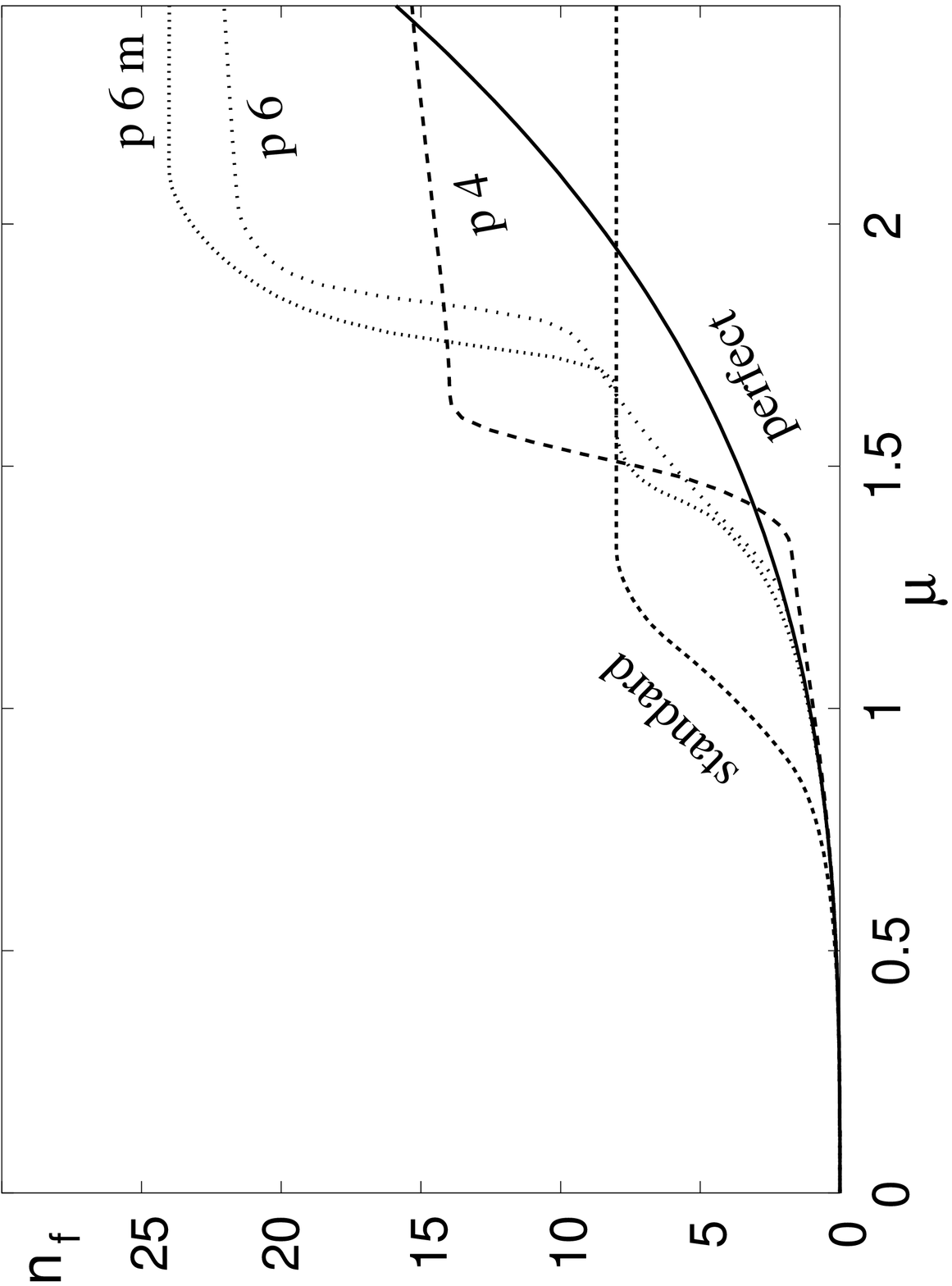}
   \end{tabular}
   \vspace{-8mm}
   \caption{{\it The fermion density for some lattice
     fermions mentioned in the text.}}
   \label{nfstag}
   \vspace{-8mm}
\end{figure}

We also show these curves for various types of staggered
fermions. The standard bound is 8, but if couplings
over temporal distance $3$ are involved, then this bound is
amplified by a factor of 3. Here we also show some Symanzik
improved actions with diagonal couplings from Ref. \cite{Bielef}.
One type called p4 has maximal temporal coupling distance 2 and
saturates at $n_{f}=16$, the others couple over distance 3 like
the truncated perfect fermions. Also here the saturation is
delayed, and we mention that especially the p6 action is doing
very well in the scaling tests too.

\vspsec
\section{Generalizations}
\vspsec

A number of {\em generalizations} are straightforward, for instance
the steps to finite temperature and finite mass.
The latter has been considered in Ref. \cite{alt} for naive
fermions, and the behavior is similar in all other cases: 
the saturation is delayed
as $m$ grows, but the upper bound is not affected.

It is fashionable to use anisotropic lattices
in thermodynamic simulations. In particular,
$\bar \xi = a_{spatial}/a_{temporal} >1$
is useful, because it provides a better resolution of
a decay in the temporal direction.
Remarkably, the couplings in a perfect action are independent
of $\bar \xi$. This property still holds after truncation by
periodic boundary conditions.
In the previous plots, we would reproduce the same curves, and the
horizontal axes now means $a_{temporal}\cdot \mu$. However,
the lattice units are defined by $a_{spatial}$,
so $\mu$ is multiplied by $\bar \xi$.\\

\vspsec
\section{Conclusions}
\vspsec

So far, $\mu$ has only been included in standard lattice actions.
We have seen how to do this in general. Of interest are applications
to improved actions, and there the improved quality 
(perfectness, classical perfectness, or Symanzik improvement 
to some order) is preserved under the inclusion of $\mu$.

We have shown that an improved action allows us to go beyond
the standard limits for the baryon resp. fermion density
(in spite of a widespread believe that this is impossible).

Applicable truncated perfect actions extend the scaling region
of a free fermion gas by one order of magnitude, which suggests
that simulations with them can be carried out on very coarse
lattices. This is not a direct remedy of the sign problem mentioned
in the introduction (because that is a problem of statistics, whereas
the improved action cures systematic errors), but it helps
indirectly, because larger physical volumes can be handled.

For the fermion-gauge couplings and for the pure gauge part,
there is much work going on to construct Symanzik improved or
approximately perfect
actions \cite{Lat97,perfrev}. If this is achieved
at $\mu =0$, then the extension to
$\mu \neq 0$ is solved by the prescription shown here.

Hopefully this is a step towards more conclusive simulations
for $n_{B} \neq 0$.\\

\vspace*{-1mm}
{\it
Most of this talk is based on Ref. \cite{chempap}; I am indebted
to U.-J. Wiese for his collaboration. I also thank M.-P. Lombardo
and F. Karsch, both, for organizing this workshop and for instructive
discussions.}

\end{document}